\documentclass[conference]{IEEEtran}
\usepackage{amsmath}
\usepackage{graphicx}
\usepackage{cite}

 \usepackage{amsfonts}
\usepackage{amsmath}
\usepackage{algpseudocode}
\usepackage{algorithm}
 \usepackage{multirow}
\algrenewcommand\algorithmicrequire{\textbf{Require:}}
\usepackage{algpseudocode}
\usepackage{amsmath}
\usepackage{subcaption} % Add this in your preamble
\usepackage{amsmath, amssymb, graphicx}
\usepackage{booktabs}
\usepackage{hyperref}
\usepackage[table]{xcolor}
\usepackage[graphicx]{realboxes}
\usepackage{titlesec}
\usepackage{pifont}% http://ctan.org/pkg/pifont
\usepackage{amssymb}% http://ctan.org/pkg/amssymb
\makeatletter
\makeatletter
\def\ps@IEEEtitlepagestyle{%
  \def\@oddfoot{\mycopyrightnotice}%
  \def\@evenfoot{}%
}
\def\mycopyrightnotice{%
  {\footnotesize 
  }% <--- Change here
  \gdef\mycopyrightnotice{}% just in case
}

\makeatother
\title{Adversarial Augmentation and Active Sampling for Robust Cyber Anomaly Detection.
}
\author{
  \IEEEauthorblockN{
      Sidahmed Benabderrahmane%\IEEEauthorrefmark{1}, 
      ,
        Talal Rahwan%\IEEEauthorrefmark{1}
 } \vspace{-1mm}\\
[10pt]
  \IEEEauthorblockA{%\IEEEauthorrefmark{1} 
  New York University, NYUAD, Division of Science.} \\
    \IEEEauthorblockA{\{sidahmed.benabderrahmane\}@gmail.com} \\
}

\usepackage[justification=raggedright,singlelinecheck=false]{caption}

\begin{document}

\maketitle

\begin{abstract}

Advanced Persistent Threats (APTs) present a considerable challenge to cybersecurity due to their stealthy, long-duration nature. Traditional supervised learning methods typically require large amounts of labeled data, which is often scarce in real-world scenarios. This paper introduces a novel approach that combines AutoEncoders for anomaly detection with active learning to iteratively enhance APT detection. By selectively querying an oracle for labels on uncertain or ambiguous samples, our method reduces labeling costs while improving detection accuracy, enabling the model to effectively learn with minimal data and reduce reliance on extensive manual labeling. We present a comprehensive formulation of the Attention Adversarial Dual AutoEncoder-based anomaly detection framework and demonstrate how the active learning loop progressively enhances the model's performance. The framework is evaluated on real-world, imbalanced provenance trace data from the DARPA Transparent Computing program, where APT-like attacks account for just 0.004\% of the data. The datasets, which cover multiple operating systems including Android, Linux, BSD, and Windows, are tested in two attack scenarios. The results show substantial improvements in detection rates during active learning, outperforming existing methods. 
\end{abstract}
\begin{IEEEkeywords}
Anomaly Detection, Deep Learning, Similarity Search, AutoEncoders, Advanced Persistent Threats.

\end{IEEEkeywords}

\section{Introduction and Background}

Advanced Persistent Threats (APTs) are among the most sophisticated and dangerous forms of cyberattacks, targeting critical infrastructure, government entities, and private data. These attacks are characterized by their stealth, persistence, and technical complexity, employing tactics such as social engineering, zero-day vulnerabilities, and lateral movement to remain undetected within networks for extended periods. APTs are particularly challenging for traditional cybersecurity tools, such as antivirus software and Intrusion Detection Systems (IDS), which rely on predefined signatures or patterns and are ill-equipped to handle the stealthy and evolving nature of these threats \cite{ghafir2014advanced,shackelford2016protecting}.\\
Unlike opportunistic attacks, APTs operate over distinct phases: reconnaissance, initial exploitation, establishing persistence, lateral movement, and data exfiltration. Their ability to blend into normal activities, such as using legitimate credentials or employing advanced obfuscation techniques, further complicates detection efforts \cite{SALIM2023e17156,tankard2011advanced}. Notable cases like Stuxnet, Fancy Bear, Cozy Bear, and Pegasus underscore the critical risks posed by APTs and the need for adaptive detection strategies. For instance, Pegasus demonstrated the severity of zero-click exploits by allowing attackers to access sensitive device data without user interaction \cite{Stuxnet11,marczak2018hide,saad2020attribution}.\\

Detection of APTs faces multiple challenges. Their stealthy operations and low signal-to-noise ratio make them indistinguishable from legitimate activities such as software updates or administrative tasks. Anomalies, which serve as indicators of malicious intent, often represent only a small fraction of real-world data, making APT detection reliant on scarce labeled samples. Furthermore, anomaly detection methods often produce high false-positive rates, confusing benign activities with genuine threats, which can overwhelm Security Operations Centers (SOCs) with irrelevant alerts \cite{han2018tapp,jenkinson2017applying}.

To overcome these challenges, our work introduces FLAGUS, a competitive framework that enhances anomaly detection capabilities for APT scenarios by combining Feedback Learning, Generative Adversarial Networks (GANs), and an Attention-based Adversarial Dual AutoEncoder (ADAEN). This integrated system addresses several limitations of traditional approaches, leveraging advanced components to improve APT detection rates while minimizing the reliance on labeled data.

Firstly, the proposed ADAEN AutoEncoder learns compact representations of normal behavior, flagging anomalies through high reconstruction errors. By incorporating attention mechanisms, ADAEN identifies critical features while adversarial training ensures robust feature extraction. After that, Active learning (AL)  strengthens FLAGUS by iteratively querying an oracle to label the most uncertain samples, significantly reducing labeling costs and improving accuracy over successive iterations \cite{hsu2015active}. Additionally, FLAGUS utilizes GANs to generate realistic synthetic data, enriching the dataset and addressing imbalances, a critical issue in APT datasets where attacks represent as little as 0.004\% of the data \cite{creswell2018generative}.

Existing anomaly detection approaches for APTs vary in methodology. Techniques like Attribute Value Frequency (AVF) assign rarity scores to attributes \cite{koufakou_2007}, One-Class Classification by Compression (OC3) leverages data compression principles \cite{smets2011}, and methods such as VF-ARM and VR-ARM detect violations in frequent and rare association rules, respectively \cite{BerradaCBMMTW20}. While effective in some scenarios, these methods suffer from key limitations, including imbalanced datasets, high false-positive rates, and inadequate correlation analysis. \\
Many publicly available datasets, such as KDDCUP99 and CICIDS2017, do not reflect the complex, stealthy nature of APTs, focusing instead on conventional attacks like denial-of-service \cite{stiawan2020cicids,khraisat2019survey,mchugh2000testing,AHMED201619net}.
Our approach leverages the DARPA ADAPT dataset, part of the Transparent Computing program, which provides rich provenance data across multiple operating systems \cite{BerradaCBMMTW20,Benabderrahmane21}. This dataset enables the modeling of subtle, correlated activities that often characterize APT attacks. By combining provenance data with a novel anomaly detection technique, FLAGUS addresses structural challenges, improves detection accuracy, and provides a scalable, adaptable framework for real-world APT detection. Evaluations over multiple operating systems demonstrate FLAGUS's superior performance over nine benchmark methods, highlighting its potential as an efficient and robust solution for detecting anomalies in cybersecurity contexts. The code and data are made available for research reproducibility.
%%%%%%%%%%%%%%%%%%%%%%%%y66666666
\section{FLAGUS: A Feedback Learning Adversarial Dual AutoEncoder for APT Detection }
\subsection{Overview:}
\begin{figure*}
\centering
%\vspace*{-25mm}
\includegraphics[width=0.8\linewidth,height=\textheight]{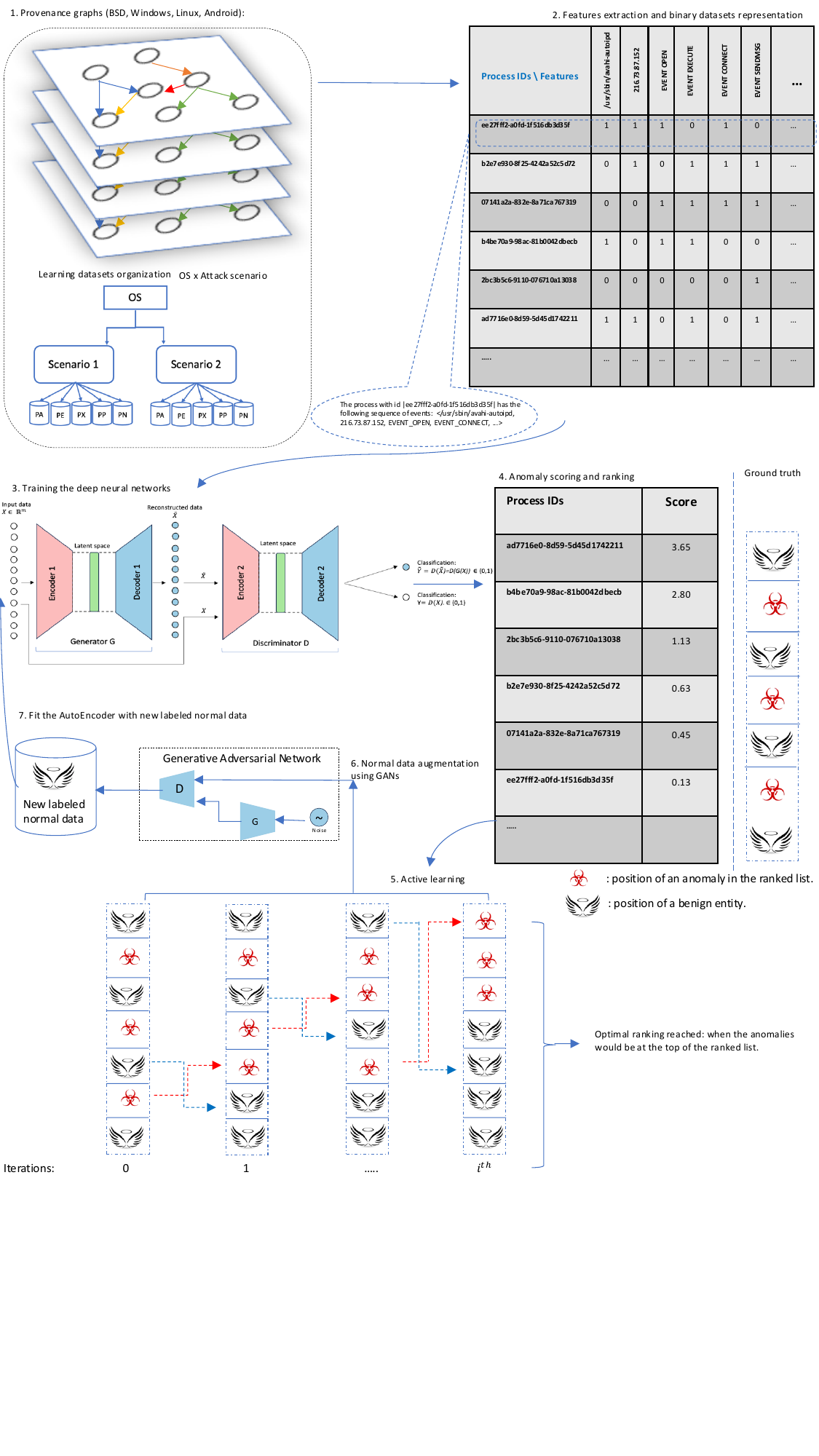}
\vspace{-32mm}
     \caption{General workflow of FLAGUS APT detection system.
      Firstly, the model is trained to learn the patterns of normal behavior from a provenance graph. Anomalies are detected based on reconstruction errors, and ranked for review. Active learning is applied iteratively, where uncertain samples are queried and labeled by an oracle. The labeled data is then augmented with a GAN and used to retrain and refine the model, improving the accuracy of the anomaly detection system over time.
     }\label{Fig:FLAGUS}
    
\end{figure*}

The overall architecture of \textit{FLAGUS} framework is shown in Figure \ref{Fig:FLAGUS} and iteratively explained in Algorithm 1.
%%%%%%%%%%%%%%%%%%%%%%%%
%%%%%%%%%%%%%%%%%%%%%%%%
It consists of three primary components. The first handles data formatting and preparation. The second, referred to as \textit{ADAEN}, is responsible for training the neural network. The third component, named \textit{AL}, focuses on active learning and GAN-based augmentation of normal labeled data.

The workflow, illustrated in Figure \ref{Fig:FLAGUS}, follows seven main steps from top to bottom. It begins with feature extraction from the provenance graph databases, where event sequences are represented as binary vectors. Each data point (process) has its associated features (steps 1 and 2 in the figure).

Next, \textit{ADAEN} AutoEncoder neural network is trained using a subset of normal labeled data (step 3). The aim is for the network to learn and capture normal behavior patterns by minimizing the reconstruction error for these instances. The size of this learning subset is intentionally small to simulate the scarcity of labeled data in real-world scenarios, highlighting the main advantage of our framework: its ability to iteratively improve performance despite the limited availability of labeled data.
Once the initial training is completed, \textit{ADAEN} is tested on a separate dataset containing both normal and anomalous instances. Anomalous entities—those exhibiting deviations from normal data patterns—tend to produce higher reconstruction errors. Anomaly scores are then computed for each data point based on its reconstruction error, and the pairwise list of [data-point, score] is sorted in descending order. Entities with high errors are considered anomalous and are ranked accordingly, with higher-ranked items being more likely anomalies (step 4 in the figure). \\
The feedback learning component is integrated into \textit{FLAGUS} pipeline to reduce false positives and improve the identification of true anomalies (step 5). Rather than querying an oracle to label all data points—which can be very costly—the system selectively chooses uncertain samples from the top of the ranked list for further labeling. These are data points whose reconstruction error is near the threshold boundary, indicating the model's uncertainty in classifying them as normal or anomalous.\\
In each iteration of the active learning process:
(i) The system queries an oracle (e.g., a human expert) to label uncertain data points near the decision boundary from the ranked list.
(ii) Based on the feedback, a new subset of normal labeled data is created. Simultaneously, a synthetic dataset is generated using a GAN, which is trained to replicate the distribution of these labeled normal data. The real and synthetic datasets are merged and reintegrated into the original training set to refine the model with more normal data (step 6).
(iii) ADAEN AutoEncoder is then retrained on the updated dataset, enhancing its ability to differentiate between normal and anomalous processes (step 7).\\
The goal is to rapidly achieve an optimal ranking, where anomalies consistently appear at the top of the ranked list after just a few iterations. Through active learning and feedback analysis, the model improves the accuracy of the rankings by incorporating new labels. This iterative process continues until the model reaches a stable state, where further iterations lead to minimal changes in the ranking. Once this equilibrium is reached, the model is considered optimized for anomaly detection, having developed a more precise understanding of both normal and anomalous behaviors.
%%%%%
\begin{algorithm}
\label{alg:flagus}
\caption{FLAGUS: Feedback-Driven Learning with GAN-Augmented Data and Uncertainty Sampling}
\begin{algorithmic}[1]
\Require Unlabeled dataset $\mathcal{U}$, initial labeled set $\mathcal{L}_0$, budget $B$, max iterations $T$
\Ensure Ranked anomaly scores for $\mathcal{U}$
\State Initialize ADAEN (Attention-based Dual Adversarial Autoencoder)
\State Train ADAEN on $\mathcal{L}_0$
\For{iteration $t = 1$ to $T$}
    \State Compute reconstruction errors for samples in $\mathcal{U}$
    \State Estimate uncertainty for each $x \in \mathcal{U}$ (e.g., using entropy or threshold margins)
    \State Select top-$k$ most uncertain samples $\mathcal{Q}_t \subset \mathcal{U}$ (query set)
    \State Augment $\mathcal{Q}_t$ using GAN to create synthetic samples $\mathcal{Q}_t^{\text{GAN}}$
    \State Obtain oracle labels for $\mathcal{Q}_t$ (if available)
    \State Update labeled set: $\mathcal{L}_{t} = \mathcal{L}_{t-1} \cup \mathcal{Q}_t \cup \mathcal{Q}_t^{\text{GAN}}$
    \State Retrain ADAEN on updated $\mathcal{L}_t$
    \State Recompute anomaly scores and update rankings on $\mathcal{U}$
    \State Evaluate nDCG on held-out test data (if available)
    \If{budget $B$ is exhausted}
        \State \textbf{break}
    \EndIf
\EndFor
\Return Ranked anomaly scores for $\mathcal{U}$
\end{algorithmic}
\end{algorithm}

%%%%
%
\subsection{ADAEN: Attention Adversarial Dual AutoEncoder }

AutoEncoders (AE) are unsupervised neural network models designed to learn lower-dimensional representations of input data, \( x \in \mathbb{R}^d \), by minimizing the reconstruction error \( L(x, \hat{x}) \). In anomaly detection, AEs are trained to capture the normal data distribution. When input data deviates from the normal data, its reconstruction error will be significantly high. This error can then be used as a detection metric, where anomaly scores are computed as: $\text{Ascore}(x) = L(x, \hat{x})$
and anomalies are flagged if \( L(x, \hat{x}) > \tau \), where \( \tau \) is a predefined threshold. The proposed ADAEN architecture (Figure \ref{Fig:FLAGUS}:3) enhances anomaly detection by combining dual AutoEncoders (AE1 and AE2) and a generative adversarial network (GAN) framework, along with an attention mechanism to improve reconstruction quality and anomaly detection.

\subsection{Dual adversarial learning:}
%\subsubsection{Dual adversarial learning:}
ADAEN comprises two AutoEncoders: the primary AutoEncoder (AE1) and the secondary AutoEncoder (AE2). Each AutoEncoder learns a different representation of the data, enhancing robustness.

The primary AutoEncoder's encoding and decoding steps are defined as: $
z_1 = f_{\theta_1}(x) = \sigma(W_e x + b_e), \quad \hat{x}_1 = g_{\phi_1}(z_1) = \sigma(W_d z_1 + b_d)$.
The secondary AutoEncoder follows a similar procedure: $
z_2 = f_{\theta_2}(x) = \sigma(W_e x + b_e), \quad \hat{x}_2 = g_{\phi_2}(z_2) = \sigma(W_d z_2 + b_d)$.

The reconstruction errors of each AutoEncoder are calculated as: $
\mathcal{L}_{\text{recon}_1}(x) = \| x - \hat{x}_1 \|^2, \quad \mathcal{L}_{\text{recon}_2}(x) = \| x - \hat{x}_2 \|^2$.
The total reconstruction error for ADAEN is a weighted sum of the errors from both AutoEncoders: $
\mathcal{L}_{\text{reconstruction}} = \alpha \mathcal{L}_{\text{recon}_1}(x) + (1 - \alpha) \mathcal{L}_{\text{recon}_2}(x)$
where \( \alpha \in [0,1] \) is a hyperparameter that balances the contributions from each AutoEncoder.
%
%\textbf{Adversarial component:}

ADAEN integrates an adversarial training approach, where AE1 serves as the generator and AE2 as the discriminator. The discriminator \( D_{\psi} \) learns to distinguish between real data \( x \), and the reconstructed data \( \hat{x}_1 \) and \( \hat{x}_2 \), from AE1 and AE2 respectively. 

The adversarial loss for the discriminator is:\\
$\mathcal{L}_{D} = - \mathbb{E}_{x \sim p_{\text{data}}}[\log D_{\psi}(x)]$$-\mathbb{E}_{\hat{x}_1, \hat{x}_2 \sim p_{\text{reconstruction}}} [\log(1 - D_{\psi}(\hat{x}_1))$ 
$+ \log(1 - D_{\psi}(\hat{x}_2))]$.
The generator loss, which encourages AE1 to generate realistic reconstructions, is: \\
$
\mathcal{L}_{\text{adv}} = - \mathbb{E}_{\hat{x}_1, \hat{x}_2 \sim p_{\text{reconstruction}}}[\log D_{\psi}(\hat{x}_1) + \log D_{\psi}(\hat{x}_2)]$. 

The overall loss function combines both reconstruction and adversarial losses: $
\mathcal{L}_{\text{ADAEN}} = \mathcal{L}_{\text{reconstruction}} + \lambda \mathcal{L}_{\text{adv}}$ 
where \( \lambda \) is a hyperparameter controlling the relative importance of the adversarial loss. ADAEN improves robustness and reconstruction capabilities by leveraging both AE1 for reconstruction and AE2 to refine and balance the learning process.

\subsection{Attention mechanism:}
To enhance the generator's reconstruction (AE1), an extra \textit{attention mechanism} layer is added to the encoded features through Encoder 1 component. Its role is to help the model focus on key parts of the input data during the generation process, ensuring that important features are captured more effectively while generating synthetic data. This would help the generator create higher-quality reconstructions of normal data and better identify patterns, which is crucial when generating new synthetic data.

\subsection{Discussion:}
The dual AutoEncoder structure in ADAEN allows for the capture of different but complementary data representations, improving the system's ability to model complex patterns. The adversarial training ensures the AutoEncoders produce high-quality reconstructions, making it easier to detect anomalies by minimizing the gap between the reconstructed data and the true distribution. The attention mechanism further improves anomaly detection by focusing on crucial features during reconstruction. This combined approach enables ADAEN to achieve robust and accurate anomaly detection, leveraging both adversarial training and feature focus techniques.

\subsection{Feedback analysis: }
%\subsubsection{\textbf{Background}}
Active learning is a paradigm where the model selectively queries the user (or an oracle) to label the most informative examples, aiming to improve the model with fewer labeled examples. This approach is particularly useful in cases where labeling data is costly or time-consuming, as is the case with APTs, where labeled data is scarce. The model focuses on uncertain or difficult examples that help refine its accuracy. In cybersecurity, the motivation behind active learning for APT detection can be thought of in terms of rewards and penalties. Positive rewards are typically given for correct attack detection or successful mitigation, while penalties occur when the model incorrectly flags normal behavior as malicious, causing disruption.

\subsection{Oracle querying and ambiguous points augmentation with GAN:}
After the initial training of the \textit{ADAEN} neural network on a dataset predominantly composed of normal data, the model aims to reconstruct normal patterns with minimal error. During testing, data is passed through the AutoEncoder, reconstruction errors are computed, and anomalies are flagged by comparing these errors to a predefined threshold. To reduce reliance on extensive manual labeling, the framework employs an active learning strategy to query the most informative samples—those with reconstruction errors close to the threshold or high errors indicative of likely anomalies. The uncertainty of a sample \( x \) is measured by its proximity to the decision boundary. This uncertainty is represented mathematically by the model's confidence score: $U(x) = 1 - p(\hat{y}|x)$ where \( p(\hat{y}|x) \) is the predicted probability of the most likely class \( \hat{y} \). Samples with the highest uncertainty are selected for labeling. The framework queries an oracle (e.g., a human expert) to label these ambiguous points as normal or anomalous. Once labeled, a Generative Adversarial Network (GAN) is employed to augment the ambiguous points. Specifically, the GAN generates synthetic data that mimics the distribution of the normal labeled data. The synthetic data, combined with the labeled samples, is merged into a new augmented dataset. This dataset is then used to refine the AutoEncoder's anomaly detection threshold and retrain the model. This integrated feedback learning and data augmentation approach enhances the model's capability to detect anomalies effectively, even when labeled data is scarce.

\subsection{Iterative loop:}
The process is iterated by reapplying the AutoEncoder to the augmented dataset and querying additional samples. Over time, the model improves at distinguishing between normal and anomalous data, enhancing detection accuracy with fewer false positives and negatives. Active learning ensures an efficient labeling process, minimizing the number of samples to be labeled, which reduces costs. Several sampling strategies could be used in the active learning process. In uncertainty sampling, the model selects points near the decision boundary (i.e., near the reconstruction error threshold), where uncertainty is highest. In outlier sampling, the model selects points with the highest reconstruction errors, which are likely to be anomalous.
These two sampling strategies were preferred here, as they are more suitable for APT detection.

\subsection{Anomaly ranking and evaluation metrics:}
In APT detection, where malicious processes make up less than 0.1\% of the dataset, accuracy is misleading, as classifying all processes as benign would yield high accuracy without detecting APTs. Instead, we focus on anomaly ranking to identify malicious processes effectively. FLUGUS ranks processes by assigning higher anomaly scores to those perceived as more anomalous. This ranked approach aligns with real-world needs, where security teams prioritize reviewing high-risk items first. It is also suitable for active learning, as the model selects top-ranked samples for labeling to improve performance.
To evaluate the ranking, we use the normalized discounted cumulative gain (nDCG)~\cite{jarvelin_2002}, a metric from Information Retrieval that measures how well anomalies are ranked~\cite{BerradaCBMMTW20,Benabderrahmane21}. The nDCG score ranges from 0 to 1, with 1 indicating that all anomalies are ranked at the top.
First, the discounted cumulative gain (DCG) is computed as: $DCG = \sum_{i=1}^{N} \frac{rel_i}{\log_2{(i+1)}}$ where $rel_i$ is the relevance of the $i$\textsuperscript{th} entry. The DCG is then normalized by the ideal DCG (iDCG) to compute the normalized score: $nDCG = \frac{DCG}{iDCG}$. 
The ideal nDCG score is achieved when anomalies are ranked above normal processes, with their exact order among anomalies being irrelevant.
This ranking-based evaluation prioritizes the model’s ability to detect and rank APTs, making the nDCG score a meaningful metric for anomaly detection, especially for rare APTs. Higher nDCG values indicate that the model effectively prioritizes significant anomalies at the top ranks. This property makes nDCG particularly suitable for evaluating anomaly detection in cybersecurity scenarios, where the critical objective is not merely classification accuracy but rather ensuring the most severe and actionable anomalies are quickly identified for timely mitigation.

\section{Experimental Setup, Results, and Insights
}
The following sections detail the datasets and benchmarking protocol used to evaluate FLAGUS against nine state-of-the-art anomaly detection methods, each employing a distinct approach to detecting anomalies. The Attribute Value Frequency (AVF) \cite{koufakou_2007}, which identifies anomalies based on the rarity of attribute values, and Frequent Pattern Outlier Factor (FPOF) \cite{he_2005}, which detects outliers by identifying less frequent patterns in the data. Outlier Degree (OD) \cite{narita_2008} and One-Class Classification by Compression (OC3) \cite{smets2011} also rely on the frequency of itemsets to detect anomalies, with OD scoring potential outliers based on association rules and OC3 identifying anomalies based on poor data compression. Valid Rare Association Rule Mining (VR-ARM) complements this by identifying rare patterns that signify anomalies, whereas Valid Frequent Association Rule Mining (VF-ARM), on the other hand, focuses on frequent rule-based pattern mining to pinpoint anomalies\cite{Benabderrahmane21}. Additionally, One-Class Support Vector Machine (OC-SVM)\cite{zhang2007one} separates anomalies by learning a boundary around the normal data, while Isolation Forest (IForest) isolates anomalies by recursively partitioning data using random splits\cite{yepmo2024leveraging}. Lastly, the Elliptic Envelope (EE) method assumes the data follows a Gaussian distribution and identifies outliers as points outside the estimated envelope\cite{vishwakarma2023new}.

Initially, FLAGUS is evaluated using the complete training datasets without integrating active learning, offering a direct comparison of its core architecture against these diverse methods for detecting anomalies in complex datasets. Subsequently, the evaluation shifts to assess the impact of FLAGUS’s active learning mechanism. By iteratively querying an oracle for the most uncertain samples, FLAGUS refines its understanding of both normal and anomalous data, significantly improving detection rates while minimizing dependence on extensive labeled datasets.  
While the baseline comparisons provide insight into FLAGUS’s capabilities relative to established methods, the primary objective is to demonstrate how FLAGUS leverages active learning to achieve incremental performance gains. This approach is particularly critical in the context of Advanced Persistent Threat (APT) detection, where labeled datasets are often scarce.  
The subsequent sections outline the experimental setup, describe the datasets, and discuss the results.  

\textbf{Datasets: } 
The datasets used in this study are derived from DARPA’s Transparent Computing (TC) program~\cite{darpa}\footnote{\url{https://gitlab.com/adaptdata}}, which captures provenance data for system activities and interactions across multiple operating systems (OS). These datasets encompass system-level activities, background operations, and processes recorded during Advanced Persistent Threat (APT)-style attacks, enabling the detection of anomalous patterns by examining dependencies and interactions among system components. This comprehensive perspective helps identify behaviors that may appear benign individually but suggest malicious intent when analyzed collectively.  
The data was processed through DARPA’s ADAPT project~\cite{berrada_2019,BerradaCBMMTW20,Benabderrahmane21}, involving provenance graph ingestion, integration, and deduplication. It includes information from four OS environments—Linux, Android, Windows, and BSD—under two attack scenarios: Pandex (E1) and Bovia (E2). The datasets are structured to describe different aspects of system processes. The ProcessEvent dataset (PE) captures event types performed by processes, while ProcessExec (PX) records the executables used to start processes. ProcessParent (PP) focuses on the executables used to start parent processes, and ProcessNetflow (PN) describes the IP addresses and ports accessed by processes. Finally, ProcessAll (PA) represents a union of all attributes from the other datasets.  

Overall, these datasets consist of Boolean vectors where a value of 1 indicates the presence of a specific attribute. Combining four operating systems, two attack scenarios, and five process aspects, the study analyzes a total of forty datasets, offering a rich source of information to investigate system behaviors during APT attacks.  
%\begin{figure}
%    \centering
%    \includegraphics[width=0.7\linewidth]{data3.png}
%    \caption{Organization of the DARPA's TC datasets. Each OS undergoes two attack scenarios, each of which contains five datasets. With four OS (BSD, Windows, Linux, Android), two attack scenarios, and five aspects (PE, PX, PP, PN, PA), a total of forty individual datasets are composed.}
%    \label{fig:enter-label}
%\end{figure}
They are described in Table~\ref{datatable} whereby the last column provides the number of attacks in each dataset. The substantially imbalanced nature of the datasets is clearly seen here. The experiments were conducted on a machine running macOS 14.5 with an Apple M1 silicon chip, 64 GB RAM. The anomaly detection models were tested on every one of the 40 datasets.
\begin{table*}[!th]
%\vspace{-6 em}
\centering
\small
\resizebox{0.99\textwidth}{!}{
\begin{tabular}{|l|l||l|l|l|l|l|l|l|l|}
\hline & Scenario & Size& $PE$   & $PX$  & $PP$  & $PN$     & $PA$  & $nb\_attacks$    & $\%\frac{nb\_attacks}{nb\_processes}$     \\ \hline \hline
BSD    & 1 &288 MB &76903 / 29  & 76698 / 107  & 76455 / 24  & 31 / 136  & 76903 / 296 & 13&0.02\\  
    & 2 &1.27 GB &224624 / 31  &224246 / 135  & 223780 / 37  & 42888 / 62 &  224624 / 265      & 11&0.004\\ \hline
Windows & 1 &743 MB & 17569 / 22    &  17552 / 215  &   14007 / 77        &   92 / 13963      & 17569 / 14431& 8&0.04\\  
   & 2 &9.53 GB& 11151 / 30    &  11077 / 388  & 10922 / 84  & 329 / 125      &  11151 / 606    &8&0.07\\ \hline
Linux  & 1 &2858 MB &247160 / 24 & 186726 / 154 & 173211 / 40 & 3125 / 81 & 247160 / 299  &25&0.01\\
    & 2 &25.9 GB &282087 / 25 & 271088 / 140 & 263730 / 45 &6589 / 6225 &  282104 / 6435      &46&0.01\\ \hline
Android& 1 &2688 MB&102 / 21     &102 / 42&0 / 0&8 / 17& 102 / 80&9&8.8\\
&2 &10.9 GB&12106 / 27     &12106 / 44&0 / 0&4550 / 213&12106 / 295 &13&0.10\\ \hline
\end{tabular}
}

\caption{Experimental datasets of DARPA's TC program used in our study. A dataset entry (columns 4 to 8) is described by a number of rows (processes) / number of columns (attributes). For instance, with ProcessAll (PA) obtained from the second scenario using Linux, the dataset has 282104 rows and 6435 attributes with 46 APT attacks (0.01\%). }
 \label{datatable}
 %\vspace*{-2\baselineskip}
\end{table*}
%%%%%%%%%%%%%%%%%%%%%%%%%%%%%%%%%%
%\subsubsection{Architecture}
%\subsubsection{Computational environment}
% Further analyses of the models, as introduced in the following sections, were conducted on the \verb|ProcessAll| dataset under the Pandex scenario recorded on a Linux system. This dataset consists of 299 features and 247160 data points. Importantly, only 25 data points (less than 0.009\% of the dataset) are anomalous.
\subsection{Results}

%\subsubsection{Comparison with existing methods}
\textbf{Comparison with existing methods:}
The nDCG scores in Table \ref{tab:ndcgscores} show that FLAGUS consistently achieves the highest or near-highest nDCG scores across all operating systems and attack scenarios, excelling in complex cases like the Bovia scenario on BSD and Linux. 
This consistency across highly heterogeneous forensic configurations indicates the robustness of FLAGUS to both platform-specific noise and attacker variability — a key requirement for real-world deployment in diverse environments such as enterprise IT or industrial control systems. For example, in BSD Bovia (PP configuration), FLAGUS achieves 0.98, far outperforming competitors like OC3, AVF, FPOF, OC-SVM, IForest and EE, which fail to exceed 0.50.  
VR-ARM proves competitive in Windows and Android environments, occasionally matching FLAGUS. Notable examples include nDCG scores of 0.82 in Windows $\times$ Pandex $\times$ PE and 0.87 in Android $\times$ Pandex $\times$ PE. While VR-ARM performs well in certain configurations, it lacks the consistency seen in FLAGUS across all OS-attack combinations. FLAGUS’s feedback loop and attention-guided latent modeling appear to generalize better in high-dimensional and under-sampled regimes.
AVF performs reasonably well in simpler cases, such as Android Pandex (PE, 0.84), but struggles in complex scenarios. Similarly, OC3 achieves decent results in a few cases, such as Android Pandex (PA, 0.82), but often fails in high-complexity datasets. It is worth noting that methods like AVF and OC3 occasionally match FLAGUS in simpler, low-dimensional scenarios, suggesting that these classical models still offer value in constrained or low-complexity cases. However, their sharp performance drop in high-dimensional, adversarial environments reinforces the need for adaptive, feedback-driven models like FLAGUS.
OD (Outlier Degree) and FPOF (Frequent Pattern Outlier Factor) perform poorly, frequently yielding DNF results and struggling across all configurations. OC-SVM, IForest, and EE also do not match FLAGUS, particularly in challenging, high-dimensional environments. The FLAGUS broad superiority, particularly in the complex Bovia and Pandex scenarios, highlights its strength in settings where traditional methods often fail (DNF or near-zero nDCG), and positions it as a reliable detector across operational threat models.
%====== 
Colored values in the same table summarize these findings with an emphasis on the highest nDCG scores and the corresponding winner methods. For each OS/attack scenario, the max nDCG value is highlighted in red. Observe that FLAGUS performed best among competitors in 7 out of 8 forensic configurations and demonstrated superior performance, achieving the highest scores in multiple datasets and scenarios.
\begin{table*}[]
\scriptsize
\centering
\begin{tabular}{lccccccc}
\hline
\textbf{Operating System} & \textbf{Attack Scenario} & \textbf{Algorithm}                                             & \textbf{PA}                                                  & \textbf{PE}                                                  & \textbf{PX} & \textbf{PP}                                                  & \textbf{PN}                                                  \\ \hline
Cadets (BSD)              &                          & \cellcolor[HTML]{FFFC9E}{\color[HTML]{FE0000} \textbf{FLAGUS}} & \cellcolor[HTML]{FFFC9E}{\color[HTML]{FE0000} \textbf{0.88}} & 0.75                                                         & 0.78        & 0.65                                                         & 0.35                                                         \\
                          &                          & AVF                                                            & 0.52                                                         & 0.51                                                         & 0.34        & 0.21                                                         & 0.58                                                         \\
                          &                          & OC3                                                            & 0.38                                                         & 0.43                                                         & 0.49        & 0.43                                                         & 0.24                                                         \\
                          &                          & OD                                                             & 0.19                                                         & 0.19                                                         & 0.15        & 0.13                                                         & 0.14                                                         \\
                          &                          & FPOF                                                           & 0.21                                                         & 0.2                                                          & 0.15        & 0.13                                                         & 0.13                                                         \\
                          &                          & VR-ARM                                                         & 0.36                                                         & 0.64                                                         & 0.08        & 0.29                                                         & 0.58                                                         \\
                          & \multirow{-7}{*}{Pandex} & VF-ARM                                                         & 0.18                                                         & 0.33                                                         & 0.53        & 0.67                                                         & 0.11                                                         \\
                          & \multicolumn{1}{l}{}     & OC-SVM                                                         & 0.13                                                         & 0.14                                                         & 0.14        & 0.11                                                         & 0.10                                                         \\
                          & \multicolumn{1}{l}{}     & IForest                                                        & 0.13                                                         & 0.14                                                         & 0.58        & 0.11                                                         & 0.25                                                         \\
                          & \multicolumn{1}{l}{}     & EE                                                             & 0.22                                                         & 0.15                                                         & 0.15        & 0.11                                                         & 0.15                                                         \\ \cline{2-8} 
                          & Bovia                    & \cellcolor[HTML]{FFFC9E}{\color[HTML]{FE0000} \textbf{FLAGUS}} & 0.79                                                         & 0.67                                                         & 0.77        & \cellcolor[HTML]{FFFC9E}{\color[HTML]{FE0000} \textbf{0.98}} & 0.15                                                         \\
                          &                          & AVF                                                            & DNF                                                          & 0.19                                                         & 0.17        & 0.17                                                         & 0.18                                                         \\
                          &                          & OC3                                                            & DNF                                                          & 0.24                                                         & 0.51        & 0.29                                                         & 0.50                                                         \\
                          &                          & OD                                                             & 0.15                                                         & 0.17                                                         & 0.17        & 0.09                                                         & 0.20                                                         \\
                          &                          & FPOF                                                           & 0.21                                                         & 0.13                                                         & 0.18        & 0.1                                                          & DNF                                                          \\
                          &                          & VR-ARM                                                         & 0.52                                                         & 0.12                                                         & 0.05        & 0.24                                                         & 0.6                                                          \\
                          &                          & VF-ARM                                                         & 0.14                                                         & 0.12                                                         & 0.06        & 0.06                                                         & 0.18                                                         \\
                          & \multicolumn{1}{l}{}     & OC-SVM                                                         & DNF                                                          & 0.12                                                         & 0.13        & 0.10                                                         & 0.12                                                         \\
                          & \multicolumn{1}{l}{}     & IForest                                                        & DNF                                                          & 0.12                                                         & 0.22        & 0.10                                                         & 0.12                                                         \\
                          & \multicolumn{1}{l}{}     & EE                                                             & 0.01                                                         & 0.12                                                         & 0.12        & 0.08                                                         & 0.15                                                         \\ \hline
5dir (Windows)            & Pandex                   & \cellcolor[HTML]{FFFC9E}{\color[HTML]{FE0000} \textbf{FLAGUS}} & 0.72                                                         & \cellcolor[HTML]{FFFC9E}{\color[HTML]{FE0000} \textbf{0.82}} & 0.24        & 0.17                                                         & 0.65                                                         \\
                          &                          & AVF                                                            & 0.52                                                         & 0.6                                                          & 0.28        & 0.21                                                         & 0.58                                                         \\
                          &                          & OC3                                                            & 0.49                                                         & 0.3                                                          & 0.28        & 0.21                                                         & 0.65                                                         \\
                          &                          & OD                                                             & DNF                                                          & 0.2                                                          & 0.15        & 0.1                                                          & 0.36                                                         \\
                          &                          & FPOF                                                           & DNF                                                          & 0.2                                                          & 0.15        & 0.1                                                          & 0.36                                                         \\
                          &                          & \cellcolor[HTML]{FFFC9E}{\color[HTML]{FE0000} \textbf{VR-ARM}} & 0.61                                                         & \cellcolor[HTML]{FFFC9E}{\color[HTML]{FE0000} \textbf{0.82}} & 0           & 0                                                            & 0.62                                                         \\
                          &                          & VF-ARM                                                         & 0.5                                                          & 0.33                                                         & 0           & 0                                                            & 0                                                            \\
                          & \multicolumn{1}{l}{}     & OC-SVM                                                         & DNF                                                          & 0.14                                                         & 0.17        & 0.09                                                         & 0.32                                                         \\
                          & \multicolumn{1}{l}{}     & IForest                                                        & DNF                                                          & 0.14                                                         & 0.37        & 0.15                                                         & 0.33                                                         \\
                          & \multicolumn{1}{l}{}     & EE                                                             & DNF                                                          & 0.15                                                         & 0.16        & 0.10                                                         & DNF                                                          \\ \cline{2-8} 
                          & Bovia                    & \cellcolor[HTML]{FFFC9E}{\color[HTML]{FE0000} \textbf{FLAGUS}} & \cellcolor[HTML]{FFFC9E}{\color[HTML]{FE0000} \textbf{0.45}} & 0.31                                                         & 0.43        & 0.43                                                         & 0.4                                                          \\
                          &                          & AVF                                                            & DNF                                                          & 0.21                                                         & 0.22        & 0.22                                                         & 0.18                                                         \\
                          &                          & OC3                                                            & DNF                                                          & 0.23                                                         & 0.24        & 0.22                                                         & 0.24                                                         \\
                          &                          & OD                                                             & DNF                                                          & DNF                                                          & DNF         & DNF                                                          & DNF                                                          \\
                          &                          & FPOF                                                           & DNF                                                          & DNF                                                          & DNF         & DNF                                                          & DNF                                                          \\
                          &                          & VR-ARM                                                         & 0.35                                                         & 0.19                                                         & 0           & 0                                                            & 0.3                                                          \\
                          &                          & VF-ARM                                                         & 0.07                                                         & 0.13                                                         & 0           & 0                                                            & 0                                                            \\
                          & \multicolumn{1}{l}{}     & OC-SVM                                                         & DNF                                                          & 0.16                                                         & 0.15        & 0.30                                                         & 0.11                                                         \\
                          & \multicolumn{1}{l}{}     & IForest                                                        & DNF                                                          & 0.16                                                         & 0.27        & 0.17                                                         & 0.11                                                         \\
                          & \multicolumn{1}{l}{}     & EE                                                             & 0                                                            & 0.16                                                         & 0.23        & 0.16                                                         & 0                                                            \\ \hline
Trace (Linux)             & Pandex                   & \cellcolor[HTML]{FFFC9E}{\color[HTML]{FE0000} \textbf{FLAGUS}} & \cellcolor[HTML]{FFFC9E}{\color[HTML]{FE0000} \textbf{0.77}} & 0.65                                                         & 0.36        & 0.23                                                         & 0.45                                                         \\
                          &                          & AVF                                                            & 0.29                                                         & 0.27                                                         & 0.43        & 0.2                                                          & 0.31                                                         \\
                          &                          & OC3                                                            & 0.41                                                         & 0.38                                                         & 0.3         & 0.24                                                         & 0.38                                                         \\
                          &                          & OD                                                             & 0.18                                                         & 0.18                                                         & 0.18        & 0.17                                                         & 0.23                                                         \\
                          &                          & FPOF                                                           & 0.18                                                         & 0.18                                                         & 0.18        & 0.17                                                         & 0.23                                                         \\
                          &                          & VR-ARM                                                         & 0.54                                                         & 0.13                                                         & 0.12        & 0                                                            & 0.58                                                         \\
                          &                          & VF-ARM                                                         & 0.13                                                         & 0.22                                                         & 0.1         & 0.12                                                         & 0.42                                                         \\
                          & \multicolumn{1}{l}{}     & OC-SVM                                                         & 0.17                                                         & 0.17                                                         & 0.18        & 0.21                                                         & 0.32                                                         \\
                          & \multicolumn{1}{l}{}     & IForest                                                        & 0.21                                                         & 0.17                                                         & 0.27        & 0.17                                                         & 0.21                                                         \\
                          & \multicolumn{1}{l}{}     & EE                                                             & DNF                                                          & 0.17                                                         & DNF         & 0.17                                                         & DNF                                                          \\ \cline{2-8} 
                          & Bovia                    & \cellcolor[HTML]{FFFC9E}{\color[HTML]{FE0000} \textbf{FLAGUS}} & \cellcolor[HTML]{FFFC9E}{\color[HTML]{FE0000} \textbf{0.87}} & 0.56                                                         & 0.38        & 0.47                                                         & 0.77                                                         \\
                          &                          & AVF                                                            & DNF                                                          & 0.29                                                         & 0.42        & 0.25                                                         & 0.42                                                         \\
                          &                          & OC3                                                            & DNF                                                          & 0.38                                                         & 0.42        & 0.42                                                         & 0.35                                                         \\
                          &                          & OD                                                             & DNF                                                          & 0.21                                                         & 0.2         & 0.2                                                          & 0.23                                                         \\
                          &                          & FPOF                                                           & DNF                                                          & 0.22                                                         & 0.2         & 0.2                                                          & 0.31                                                         \\
                          &                          & VR-ARM                                                         & 0.45                                                         & 0.14                                                         & 0           & 0                                                            & 0.39                                                         \\
                          &                          & VF-ARM                                                         & 0.09                                                         & 0.1                                                          & 0.004       & 0.03                                                         & 0.11                                                         \\
                          & \multicolumn{1}{l}{}     & OC-SVM                                                         & DNF                                                          & 0.18                                                         & 0.17        & 0.19                                                         & 0.36                                                         \\
                          & \multicolumn{1}{l}{}     & IForest                                                        & DNF                                                          & 0.18                                                         & 0.33        & 0.32                                                         & 0.20                                                         \\
                          & \multicolumn{1}{l}{}     & EE                                                             & 0                                                            & 0.18                                                         & 0.22        & 0.18                                                         & 0                                                            \\ \hline
Clearscope (Android)      & Pandex                   & \cellcolor[HTML]{FFFC9E}{\color[HTML]{FE0000} \textbf{FLAGUS}} & 0.80                                                         & \cellcolor[HTML]{FFFC9E}{\color[HTML]{FE0000} \textbf{0.87}} & 0.45        & NA                                                           & 0.68                                                         \\
                          &                          & AVF                                                            & 0.83                                                         & 0.84                                                         & 0.39        & NA                                                           & 0.47                                                         \\
                          &                          & OC3                                                            & 0.82                                                         & 0.74                                                         & 0.39        & NA                                                           & 0.64                                                         \\
                          &                          & OD                                                             & 0.34                                                         & 0.33                                                         & 0.22        & NA                                                           & 0.36                                                         \\
                          &                          & FPOF                                                           & 0.31                                                         & 0.29                                                         & 0.22        & NA                                                           & 0.42                                                         \\
                          &                          & \cellcolor[HTML]{FFFC9E}{\color[HTML]{FE0000} \textbf{VR-ARM}} & 0                                                            & \cellcolor[HTML]{FFFC9E}{\color[HTML]{FE0000} \textbf{0.87}} & 0           & NA                                                           & 0.46                                                         \\
                          &                          & VF-ARM                                                         & 0                                                            & 0.77                                                         & 0           & NA                                                           & 0                                                            \\
                          & \multicolumn{1}{l}{}     & OC-SVM                                                         & 0.32                                                         & 0.44                                                         & 0.37        & NA                                                           & 0.50                                                         \\
                          & \multicolumn{1}{l}{}     & IForest                                                        & 0.34                                                         & 0.32                                                         & 0.78        & NA                                                           & 0.68                                                         \\
                          & \multicolumn{1}{l}{}     & EE                                                             & 0.4                                                          & 0.32                                                         & 0.54        & NA                                                           & 0.54                                                         \\ \cline{2-8} 
                          & Bovia                    & FLAGUS                                                         & 0.52                                                         & 0.67                                                         & 0.51        & NA                                                           & 0.41                                                         \\
                          &                          & AVF                                                            & 0.35                                                         & 0.3                                                          & 0.38        & NA                                                           & 0.32                                                         \\
                          &                          & OC3                                                            & 0.40                                                         & 0.32                                                         & 0.39        & NA                                                           & 0.30                                                         \\
                          &                          & OD                                                             & 0.20                                                         & 0.22                                                         & 0.29        & NA                                                           & 0.34                                                         \\
                          &                          & FPOF                                                           & 0.37                                                         & 0.36                                                         & 0.29        & NA                                                           & 0.36                                                         \\
                          &                          & \cellcolor[HTML]{FFFC9E}{\color[HTML]{FE0000} \textbf{VR-ARM}} & 0.51                                                         & 0.5                                                          & 0           & NA                                                           & \cellcolor[HTML]{FFFC9E}{\color[HTML]{FE0000} \textbf{0.71}} \\
                          &                          & VF-ARM                                                         & 0.43                                                         & 0.12                                                         & 0           & NA                                                           & 0.1                                                          \\
                          & \multicolumn{1}{l}{}     & OC-SVM                                                         & 0.17                                                         & 0.18                                                         & 0.17        & NA                                                           & 0.16                                                         \\
                          & \multicolumn{1}{l}{}     & IForest                                                        & 0.22                                                         & 0.18                                                         & 0.49        & NA                                                           & 0.23                                                         \\
                          & \multicolumn{1}{l}{}     & EE                                                             & 0.18                                                         & 0.19                                                         & 0.2         & NA                                                           & 0.16                                                         \\ \hline
\end{tabular}
\caption{Performance Evaluation of Anomaly Detection Methods Using nDCG Scores. Bold values represent the max nDCG for each each forensic configuration (OS $\times$ attack scenario $\times$ dataset) whereas red highlighted values represent the winner methods. NA: data not available. DNF: Did not finish.}
\label{tab:ndcgscores}
%\vspace*{-1\baselineskip}
\end{table*}
%==================
%%%%%%%%%%%%%%%%
%%%%%%%%%%%%%%
\textbf{Active learning assessment:}
The evaluation of FLAGUS with the active learning loop focuses on its performance across 40 datasets, emphasizing the scenarios where FLAGUS outperforms previous anomaly detection methods. 
The FLAGUS Active Learning Loop is configured to run for 40 iterations to strike a balance between model improvement and the need for timely security decision making, which is crucial in cybersecurity environments.

\begin{figure*}
% \vspace*{-6\baselineskip}
\centering
\subfloat[Reconstruction Error Histogram with Threshold. The x-axis represents reconstruction error values, and the y-axis indicates the frequency of data points for each error. Uncertain data points with high reconstruction error might represent the anomalous points that the model struggles to reconstruct.\label{fig:subim1}]{%
  \includegraphics[width=0.48\textwidth]{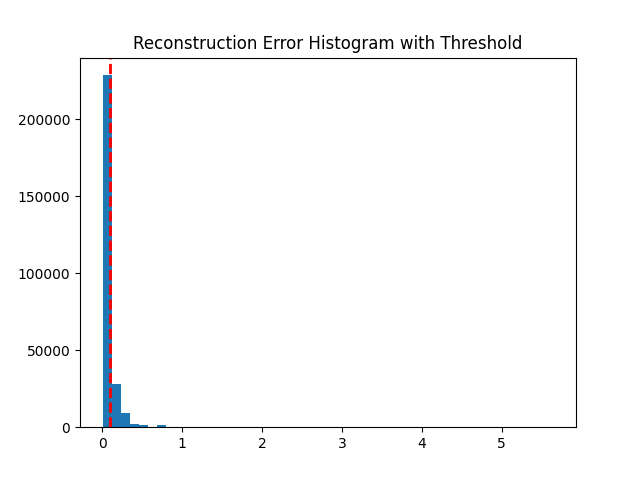}%
}\hfil
\subfloat[nDCG score variation over Active Learning iterations for the Linux PA dataset using FLAGUS (Pandex E1 scenario). The blue dashed line shows raw nDCG data, the red dashed line shows smoothed values, and a boxplot on the right shows nDCG score distribution across iterations.\label{fig:subim2}]{%
  \includegraphics[height=6cm, width=0.48\textwidth]{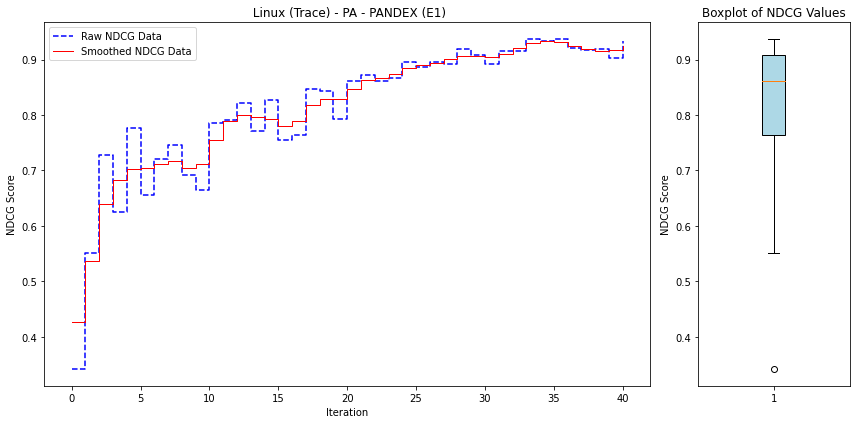}%
}
\caption{FLAGUS evaluation during active learning. (a) Reconstruction error distribution with the decision threshold. (b) Performance improvements in active learning over iterations.}
\label{fig:activelearning}
 %\vspace*{-4\baselineskip}
\end{figure*}

The reconstruction error histogram in Figure (\ref{fig:activelearning}-a) demonstrates how FLAGUS's AutoEncoder performs, with most data points having low reconstruction errors, indicating successful reconstructions. In this example data belongs to Linux$\times$PA$\times$Pandex forensic configuration (randomly chosen here for illustration purposes). Uncertain data points with high reconstruction error might represent the anomalous points that the model struggles to reconstruct, signaling potential attacks or anomalies. The anomaly detection threshold is established at the 80th percentile of the reconstruction errors of the test data set, meaning that 80\% of the data points are classified as normal, with the remaining 20\% flagged as anomalies.  This choice is flexible and can be adjusted to fine-tune the algorithm's performance. The red dashed line represents this threshold marker that is chosen to distinguish between normal and anomalous data points. Here, the uncertainty threshold can be adjusted for better performance and can be recalibrated iteratively during active learning for more accurate anomaly detection over time.  

Figure (\ref{fig:activelearning}-b) presents the evolution of the nDCG scores over several Active Learning iterations for the Linux$\times$PA$\times$Pandex forensic configuration. It consists of two subfigures: The raw nDCG data is represented by the blue dashed line showing the progression of the nDCG scores over 40 iterations, while the smoothed nDCG values are depicted by the red curve. Initially, the nDCG score starts around 0.34 (min value at cold start) but gradually improves as iterations progress, stabilizing between 0.70 and 0.94 in later iterations. The median is about 0.84 and the mean over all values is 0.82. There are noticeable fluctuations in performance, indicating varying success rates at different stages. However, the overall trend is upward, demonstrating that the model is learning and improving its ranking accuracy with each iteration. \\
Table \ref{FLAGUS-baseline-AL} summaries the performance of FLAGUS in two settings: (1) without active learning (baseline) and (2) with active learning, across different operating systems (BSD, Windows, Linux, and Android) and datasets (PA, PE, PX, PP, and PN) under two attack scenarios (Pandex and Bovia). The evaluation considers maximum, mean, and median nDCG scores to assess the model’s ranking improvement.

For BSD, active learning consistently improved nDCG scores compared to the baseline, with the maximum score often reaching 1.0, particularly in the Bovia attack scenario. Smaller datasets, like PN, showed more fluctuations, likely due to data sparsity, but active learning still resulted in overall higher nDCG scores, particularly in scenarios like PA and PE. In Windows, active learning resulted in significant improvements across most datasets and attack scenarios. The Pandex scenario achieved consistently high maximum nDCG scores, with mean and median values substantially higher than the baseline. The Bovia scenario also saw notable improvements in datasets like PA, PX, and PP, where active learning drove nDCG scores to their highest values. Although smaller datasets showed lower mean and median values, the overall trend was positive with active learning. For Linux, active learning improved performance across most datasets, with particularly notable gains in PA, PE, PX, and PN. The maximum nDCG scores were consistently higher in the active learning setting, and the variation between mean and median values was reduced, providing a more stable performance. Finally, in Android, active learning led to higher maximum and median nDCG scores, particularly in datasets like PA and PE. Both the Pandex and Bovia scenarios saw significant improvements, with the maximum nDCG scores reaching 1.0 in several instances.

Overall, the results indicate that active learning consistently enhanced performance across all operating systems and datasets, improving both consistency and peak ranking performance.

%\textcolor{red}{WE WILL ADD OTHER FIGURES AND RESULTS}

  %%%%%%%%%%%%%%%%%%%%
\begin{table*}
% \vspace*{-7\baselineskip}
\scriptsize

\centering
\begin{tabular}{lc|cccc|cccc|cccc|cccc|cccc}
\hline

\multirow{3}{*}{\textbf{\rotatebox{90}{Operating System}}} & \multirow{3}{*}{\textbf{\rotatebox{90}{Attack Scenario}}} & \multicolumn{4}{c|}{\textbf{PA}}                                                                                             & \multicolumn{4}{c|}{\textbf{PE}}                                                                                                   & \multicolumn{4}{c|}{\textbf{PX}}                                                                                                   & \multicolumn{4}{c|}{\textbf{PP}}                                                                                                   & \multicolumn{4}{c}{\textbf{PN}}                                                                                               \\
\cline{3-22} 
                                           &                                           & \multirow{2}{*}{\textbf{\rotatebox{90}{Baseline}}} & \multicolumn{3}{c|}{\textbf{\rotatebox{90}{Active Learning}}}                                           & \textbf{\rotatebox{90}{Baseline}}    & \multicolumn{3}{c|}{\textbf{\rotatebox{90}{Active Learning}}}                                                               & \textbf{\rotatebox{90}{Baseline}}    & \multicolumn{3}{c|}{\textbf{\rotatebox{90}{Active Learning}}}                                                               & \textbf{\rotatebox{90}{Baseline}}    & \multicolumn{3}{c|}{\textbf{\rotatebox{90}{Active Learning}}}                                                               & \textbf{\rotatebox{90}{Baseline}}    & \multicolumn{3}{c}{\textbf{\rotatebox{90}{Active Learning}}}                                                               \\ \cline{4-6} \cline{8-10} \cline{12-14} \cline{16-18} \cline{20-22} 
                                           &                                           &                                    & \multicolumn{1}{c}{\textbf{\rotatebox{90}{Max}}} & \textbf{\rotatebox{90}{Mean}} & \multicolumn{1}{c|}{\textbf{\rotatebox{90}{Median}}} & \multicolumn{1}{l}{} & \multicolumn{1}{c}{\textbf{\rotatebox{90}{Max}}} & \multicolumn{1}{c}{\textbf{\rotatebox{90}{Mean}}} & \multicolumn{1}{c|}{\textbf{\rotatebox{90}{Median}}} & \multicolumn{1}{l}{} & \multicolumn{1}{c}{\textbf{\rotatebox{90}{Max}}} & \multicolumn{1}{c}{\textbf{\rotatebox{90}{Mean}}} & \multicolumn{1}{c|}{\textbf{\rotatebox{90}{Median}}} & \multicolumn{1}{l}{} & \multicolumn{1}{c}{\textbf{\rotatebox{90}{Max}}} & \multicolumn{1}{c}{\textbf{\rotatebox{90}{Mean}}} & \multicolumn{1}{c|}{\textbf{\rotatebox{90}{Median}}} & \multicolumn{1}{l}{} & \multicolumn{1}{c}{\textbf{\rotatebox{90}{Max}}} & \multicolumn{1}{c}{\textbf{\rotatebox{90}{Mean}}} & \multicolumn{1}{c}{\textbf{\rotatebox{90}{Median}}} \\ 
                                           \hline
BSD                              & Pandex                                    & 0.88                                                  & 0.91                                            & 0.87                                             & 0.89                                               & 0.75                                                 & 1                                               & 0.87                                             & 0.93                                               & 0.78                                                 & 1                                               & 0.78                                             & 0.87                                               & 0.65                                                 & 0.36                                            & 0.21                                             & 0.15                                               & 0.35                                                 & 0.36                                            & 0.21                                             & 0.27                                               \\ \cline{2-22} 
                                           & Bovia                                     & 0.79                                                                  & 1                                               & 0.98                                             & 1                                                  & 0.67                                                 & 1                                               & 0.93                                             & 1                                                  & 0.77                                                 & 1                                               & 0.90                                             & 0.97                                               & 0.98                                      & 1                                               & 0.99                                             & 1                                                  & 0.15                                                 & 0.15                                            & 0.14                                             & 0.15                                               \\ \hline
Windows                             & Pandex                                    & 0.72                                                                  & 1                                               & 0.86                                             & 0.89                                               & 0.82                                        & 1                                               & 0.96                                             & 0.97                                               & 0.24                                                 & 0.79                                            & 0.56                                             & 0.58                                               & 0.17                                                 & 0.75                                            & 0.53                                             & 0.51                                               & 0.65                                                 & 0.91                                            & 0.72                                             & 0.73                                               \\ \cline{2-22} 
                                           & Bovia                                     & 0.45                                                         & 1                                               & 0.82                                             & 0.86                                               & 0.31                                                 & 0.62                                            & 0.51                                             & 0.56                                               & 0.43                                                 & 0.94                                            & 0.72                                             & 0.77                                               & 0.43                                                 & 0.97                                            & 0.78                                             & 0.75                                               & 0.4                                                  & 0.36                                            & 0.16                                             & 0.19                                               \\ \hline
Linux                            & Pandex                                    & 0.77                                                     & 0.94                                            & 0.82                                             & 0.84                                               & 0.65                                                 & 0.78                                            & 0.62                                             & 0.61                                               & 0.36                                                 & 0.69                                            & 0.39                                             & 0.36                                               & 0.23                                                 & 0.47                                            & 0.24                                             & 0.19                                               & 0.45                                                 & 0.87                                            & 0.82                                             & 0.87                                               \\ \cline{2-22} 
                                           & Bovia                                     & 0.87                                                        & 0.84                                            & 0.72                                             & 0.73                                               & 0.56                                                 & 0.62                                            & 0.56                                             & 0.56                                               & 0.38                                                 & 0.84                                            & 0.70                                             & 0.72                                               & 0.47                                                 & 0.54                                            & 0.49                                             & 0.48                                               & 0.77                                                 & 0.77                                            & 0.72                                             & 0.77                                               \\ \hline
Android                      & Pandex                                    & 0.80                                                                  & 0.99                                            & 0.84                                             & 0.89                                               & 0.87                                   & 0.99                                            & 0.87                                             & 0.96                                               & 0.45                                                 & 1                                               & 0.88                                             & 0.97                                               & NA                                                   & NA                                              & NA                                               & NA                                                 & 0.68                                                 & 0.69                                            & 0.69                                             & 0.69                                               \\ \cline{2-22} 
                                           & Bovia                                     & 0.52                                                                  & 1                                               & 0.85                                             & 0.93                                               & 0.67                                                 & 1                                               & 0.74                                             & 0.78                                               & 0.51                                                 & 1                                               & 0.85                                             & 0.96                                               & NA                                                   & NA                                              & NA                                               & NA                                                 & 0.41                                                 & 0.39                                            & 0.38                                             & 0.39                                               \\ \hline

\end{tabular}

\caption{Comparison of FLAGUS Performance with and without Active Learning (Baseline). For each dataset, the first column represents the scores where the algorithm is trained on the entire dataset from the start without any iterative data augmentation. The remaining columns (Max, Mean, Median) represent the scores when running Active Learning. The maximum captures the peak performance, the mean reflects the overall average behavior across all iterations, and the median offers a robust measure of typical performance by minimizing the impact of outliers. 
}

 %\vspace*{-2\baselineskip}
\label{FLAGUS-baseline-AL}
\end{table*}

% \usepackage{multirow}

%%%%%%%%%%%%%%%
%\begin{figure}
%    \centering
%    \includegraphics[width=1\linewidth]{ActiveLearningoutput1.png}
%    \caption{Enter Caption}
%    \label{fig:enter-label}
%\end{figure}
%%%%
%\begin{figure}
%    \centering
%    \includegraphics[width=1\linewidth]%{ActiveLearningoutput2.png}
 %   \caption{Enter Caption}
 %   \label{fig:enter-label}
%\end{figure}
%
%
%
%
%
%
\section{Conclusion}
In this paper, we presented FLAGUS, an innovative feedback-driven framework for anomaly detection in data-scarce cybersecurity environments, particularly those involving Advanced Persistent Threats (APTs). By tightly integrating an attention-guided adversarial dual autoencoder (ADAEN) with uncertainty-driven active learning and GAN-based data augmentation, FLAGUS addresses several critical challenges in real-world anomaly detection: extreme class imbalance, limited labeled data, and subtle malicious behaviors masked within benign system processes.

Our experimental results on 40 provenance-based datasets, derived from the DARPA Transparent Computing program, demonstrate that FLAGUS significantly outperforms nine state-of-the-art baseline methods across multiple OS and attack scenarios. Using nDCG ranking as the evaluation metric, we show that FLAGUS excels not only in raw anomaly detection performance, but also in effectively prioritizing high-risk events for review — a property crucial for operational use in SOC environments.

A key feature of our study is the ablation analysis enabled through the active learning setup: iteration 0 of FLAGUS (prior to any oracle feedback) serves as a natural baseline, revealing the added value introduced by each stage of the feedback loop. Through iterative querying of uncertain samples and their GAN-driven augmentation, the model consistently improves both its ranking accuracy and generalization capacity over time, while keeping labeling costs minimal.

Beyond the strong empirical performance, FLAGUS offers a scalable, architecture-agnostic approach that can be extended to other domains where labeled anomalies are rare and costly to obtain — such as fraud detection, insider threat monitoring, and medical anomaly diagnosis.

For future work, we plan to explore:
\begin{itemize}
    \item     Integrating self-supervised pretraining to further reduce dependence on labeled data,
    \item     Extending the feedback loop to incorporate reinforcement learning for policy-based querying,
    \item    And deploying FLAGUS in real-time monitoring pipelines, evaluating its latency and scalability under adversarial conditions.
\end{itemize}

Overall, FLAGUS provides a robust foundation for data-efficient anomaly detection in complex cyber environments and offers a promising direction for research at the intersection of deep learning, adversarial data generation, and human-in-the-loop AI.
\bibliographystyle{IEEEtran}
\bibliography{references2}

\end{document}